\documentclass[aps,preprint,showpacs]{revtex4}
\usepackage{epsfig}
\usepackage{amsmath}
\usepackage{amssymb}
\usepackage{amsfonts}
\usepackage{enumerate}

\begin{document}
%\draft
\title{The dominant spin relaxation mechanism in compound organic semiconductors}
\author{Supriyo Bandyopadhyay \footnote{Corresponding 
author. E-mail:
sbandy@vcu.edu}}
\affiliation{Department of Electrical and Computer Engineering, Virginia
Commonwealth University, Richmond, VA 23284, USA}
%\maketitle

\begin{abstract}

Despite the recent interest in ``organic spintronics'', the dominant spin relaxation mechanism of
electrons or holes in an organic compound semiconductor has not been conclusively
identified. There have been sporadic suggestions that it might be hyperfine interaction caused by 
background nuclear spins,
 but no confirmatory evidence to support
 this has ever been 
presented. Here, we  report the electric-field dependence of the spin diffusion
length in an organic spin-valve structure consisting of an 
$Alq_3$ spacer layer, and argue that this data, as well as available data on the 
temperature dependence of this length, contradict the notion that 
hyperfine interactions relax spin. Instead, they
suggest  that
the Elliott-Yafet mechanism, arising from spin-orbit interaction, 
 is more likely the dominant spin 
relaxing mechanism.

\end{abstract}

\pacs{ 72.25.Rb, 72.25.Dc, 72.25.Hg, 72.25.Mk}
\maketitle
\pagebreak

Spin relaxation in most solids is  caused primarily by mechanisms associated with 
spin-orbit 
and contact hyperfine interactions. Since compound organic semiconductors are typically made of light elements 
(hydrogen, oxygen and carbon), the 
spin-orbit interaction 
in them should be very weak since it is proportional to the fourth power of the 
atomic number of the constituent elements. At the same time, contact hyperfine interaction 
(between electron and nuclear spins) should  also be very weak
 -- at least in $\pi$-conjugated organic molecules -- because the $\pi$-electrons' wavefunctions
are mainly $p_z$ orbitals that have nodes in the molecular plane \cite{naber}.
As a result, organics tend to exhibit long spin relaxation times that could exceed those 
in inorganic 
semiconductors by several orders of magnitude at temperatures well
above that of liquid nitrogen. This has generated significant interest 
in organic spintronics \cite{nmat1,nmat2,nmat3,nmat4} owing to the realization that  
organic semiconductors could very well emerge as the material of choice in many spintronic 
applications.

Despite all this interest and research, a question of fundamental importance has remained unanswered: 
which of the two mechanisms -- spin-orbit interaction or hyperfine interaction -- is the dominant 
causative agent for spin relaxation in organics. Because 
spin-orbit interaction is so much weaker in organics than in inorganics, the 
trend has been to conjecture (tacitly) that contact hyperfine interaction must be the dominant 
spin relaxation mechanism \cite{nmat4,koopmans,koopmans1}. However, to our knowledge, no
conclusive evidence has ever been presented to substantiate this belief. This  remains an open question.

In this paper, we show that a large body of experimental evidence does not support the notion
that hyperfine interaction is the dominant spin relaxation mechanism in
 the most widely studied organics. Instead, it points to the Elliott-Yafet 
mechanism \cite{elliott}, arising from spin-orbit interaction, as being the more likely culprit. We reach this conclusion based on the reported 
temperature
and electric-field dependences of the spin diffusion length -- the latter reported here -- which are not 
consistent with 
hyperfine interaction, 
but are 
consistent with the Elliott-Yafet mechanism being dominant. In the rest of this paper, we elucidate the arguments 
leading to this conclusion.

At a temperature $T$ and in an electric field $E$, the spin diffusion length $L_s(E,T)$ 
of a spin carrier in any solid is related to the spin relaxation time $\tau_s(E,T)$
and the carrier mobility $\mu(E, T)$ according to the relation \cite{saikin,book}
\begin{equation}
L_s(E,T) 
= {{1}\over{ -{{e|E|}\over{2kT}}+ \sqrt{\left ({{e|E|}\over{2kT}}\right )^2 + {{e}\over{kT \mu(E,T) \tau_s(E,T)}}}}},
\end{equation}
where 
$k$ is the Boltzmann constant and $e$ is the electronic charge. 

Since carriers in organics travel by hopping from site to site assisted by thermal excitation in an electric field 
[similar to Poole Frenkel conduction] \cite{schein} the mobility is usually expressed as \cite{le}
\begin{equation}
\mu(E, T) = {{e d^2(E,T)}\over{\tau_0(E,T) \beta (T) \sqrt{E}}} tanh \left ( {{\beta (T) \sqrt{E}}\over{kT}} \right )
\exp \left ( {{\beta (T) \sqrt{E} - \Delta (T)}\over{kT}} \right ) ,
\end{equation}
where $d(E,T)$ is the mean hopping distance, $\tau_0(E,T)$ is the mean hopping time, $\Delta (T)$ is the activation 
energy for hopping and $\beta(T)$ is the field emission constant.

In low electric fields $\left ( |E| \ll \sqrt{ {{kT}\over{e \mu(E, T) \tau_s (E, T)}} } \right )$, Equation (1) simplifies to 
\begin{equation}
\left [ L_s(E,T) \right ]_{low~E}  \approx 
\sqrt{{{kT}\over{e}} \mu(0, T) \tau_s(0, T)},
\end{equation}
where (see Equation (2))
\begin{equation}
\mu(0, T) = {{e d^2(0,T)}\over{\tau_0(0,T) kT }} 
\exp \left ( {{ - \Delta (T)}\over{kT}} \right ) .
\end{equation}

Substituting Equation (4) in Equation (3), we get
\begin{equation}
\left [L_s(E,T) \right ]_{low~E} = d(0, T) \sqrt{{{\tau_s(0,T)}\over{\tau_0(0,T)}}}
\exp \left ( {{ - \Delta (T)}\over{2kT}} \right ).
\end{equation}

The last equation is very instructive. The quantities $d(0, T)$, $\tau_0(0,T)$ and $\Delta (T)$ should 
not have strong 
temperature dependence
at cryogenic temperatures. 
Therefore, 
Equation (5) 
tells us that the low-field and low-temperature spin diffusion length 
$\left [L_s(E,T) \right ]_{low~E}$ 
must increase with 
increasing temperature, {\it unless} 
$\tau_s(0,T)$ decreases with increasing temperature. This makes sense intuitively, 
and we could have predicted it from Equation (3) directly. Since carriers travel by 
Brownian motion, increased temperature should 
result in increased spin diffusion length unless the spin relaxation time decreases 
with increasing temperature. Thus, if we ever 
observe $\left [ L_s(E,T) \right ]_{low~E}$ decreasing with increasing temperature 
(at cryogenic temperatures), then we must conclude that $\tau_s(0,T)$
also decreases with rising temperature. Consequently, 
the temperature dependence of the spin diffusion length at low temperatures is very revealing; 
it tells us how the spin relaxation time varies with temperature. 
In turn, that can allow us to decipher the major spin relaxing mechanism.

So far, every experiment reported in the literature has found that in organics, 
$\left [ L_s(E,T) \right ]_{low~E}$
decreases with increasing temperature in the cryogenic range, either 
slowly \cite{pramanik_nature}, or moderately \cite{xiong}, or rapidly \cite{drew}. That then tells us that $\tau_s(0,T)$
must also decrease with rising temperature, slowly, moderately or rapidly. The rapid decrease {\it cannot} be 
consistent with 
hyperfine interaction because in that mechanism, spin relaxation is caused by the magnetic field of the nuclear 
spins, 
which, at best, can have weak temperature dependence. Indeed, theories based on spin diffusion in a disordered 
organic 
in the presence of the hyperfine magnetic field predict a weak temperature dependence of the spin diffusion 
length \cite{koopmans}.
Therefore, the observation in ref. 
\cite{drew}, 
which showed a {\it rapid} decrease of $\left [L_s(E,T) \right ]_{low~E}$ with increasing temperature, 
is clearly inconsistent with the notion that 
hyperfine interaction could have been the dominant spin relaxing mechanism in that organic.

To probe this matter further and correctly identify the dominant spin relaxation mechanism, we can investigate the 
electric-field 
dependence of the 
spin relaxation length and spin relaxation time in an organic, which, to our knowledge, has never been attempted. Here, we 
report some 
data on the 
electric-field dependence of the spin diffusion length in an organic and infer the electric-field dependence of the spin 
relaxation time 
from that data. This sheds further light on the dominant spin relaxation mechanism.

In ref. \cite{pramanik}, we carried out experiments in organic spin valves to extract the spin
diffusion length under varying electric 
fields, where 
the organic 
was tris(8-hydroxyquinolinolato aluminum) or $Alq_3$ with a chemical formula of
C$_{27}$H$_{18}$N$_3$O$_3$Al. The spin valves were nanowires of $Alq_3$ with cobalt and nickel contacts.
We will assume that $P_1$ and $P_2$ are the spin polarizations at the Fermi energy in the injecting and detecting  contacts, 
$\alpha_1$ and $\alpha_2$ are the effective spin injection and detection efficiencies at the two contacts, and  $L$ is the organic 
layer thickness.
Schottky barriers form at both contacts because of the energy level alignment \cite{align}. The picture of carrier transport
in these spin valves
presented in refs. \cite{pramanik_nature,xiong} is that carriers first tunnel through 
the Schottky barrier at the injecting 
contact with a spin polarization $P_1 \alpha_1$, then drift and diffuse through the bulk of the organic with 
exponentially decaying spin polarization $e^{ - L/L_s(E,T)}$, and finally tunnel through the
second Schottky barrier to reach the detecting contact. Therefore, the spin valve magnetoresistance ratio $\Delta R/R$ 
will be given by the modified Jullier\'e formula 
(adapted from ref. \cite{pramanik_nature,xiong}):
\begin{equation}
{{\Delta R}\over{R}} = {{2 P_1 \alpha_1 P_2 \alpha_2 e^{ - L/L_s(E,T) }}\over{1 - P_1 \alpha_1 P_2 \alpha_2 e^{ - L/L_s(E,T)}}}.
\end{equation}

In order to verify the transport picture in refs. \cite{pramanik_nature,xiong}, we had measured the current-voltage (I-V) characteristics of the spin valves 
at varying temperatures \cite{pramanik_nature}. They
were nearly temperature-independent and almost piecewise linear.  The current increased 
quasi-linearly with a small slope up to a 
threshold voltage of $\sim$ 2 V, and then increased rapidly with a much larger slope. 
This behavior is inconsistent with tunneling through pinholes in the organic --
which has been proposed as an alternate transport model \cite{argonne} -- since that would have produced two 
 features which are absent. First, tunneling causes the I-V characteristic to be superlinear but smooth (not abrupt like a piecewise
 linear characteristic) \cite{szc1}, and second, tunneling makes the junction resistance temperature-dependent \cite{szc1}. 
 Since neither
 feature is observed, we can rule out tunneling through pinholes. Refs. \cite{santos,szc2} carried out high resolution 
 transmission electron microscopy studies of ferromagnet/organic junctions and found them to be abrupt with no evidence of 
 interdiffusion. This further eliminates the existence of pinholes in the organic since they would have 
 caused interdiffusion. 
 
 The 
 observed I-V behavior is however very consistent with the transport picture presented in 
 refs. \cite{pramanik_nature,xiong}. At low bias voltages, the Schottky barrier at the 
 injecting contact is thick enough to suppress tunneling so that the current is 
 mostly due to thermionic emission. With increasing bias, the interface Schottky barrier becomes progressively thinner owing to 
 band bending and the tunneling increases. At some threshold bias, the tunneling current 
 exceeds that  due to thermionic emission. Thereafter, the tunneling injection dominates and with increasing 
 bias (decreasing barrier thickness) it increases rapidly. Thus, the current  remains small up 
 to a threshold bias ($\sim$ 2V), at which point cross over from thermionic emission to tunneling takes place,
 and then the current takes off. Since the tunneling probability is independent of temperature, the I-V characteristic is 
 virtually temperature-independent. Therefore, the observed I-V characteristic is in qualitative 
 agreement with the 
 transport picture presented in refs. \cite{pramanik_nature,xiong}.

In ref. \cite{pramanik}, we measured the ratio $\Delta R/R$ as a 
function of 
electric current through the organic at a temperature of 1.9 K \cite{pramanik}. 
In our samples, $L$  $\approx$ 30 nm. We assume $P_1$ = 0.4 (cobalt contact) \cite{tsymbal}, $P_2$ = 0.3 (nickel contact) \cite{tsymbal}, 
and, $\alpha_1 = \alpha_2 = 1$, which is the same assumption as in ref. \cite{pramanik_nature,xiong}. In reality,
the spin injection and detection efficiencies are never quite 100\%, but they can be very high at organic/ferromagnet
interfaces -- as high as 85-90\% in some cases \cite{cinchetti}. Since 
$L_s(E,T)$ is not very sensitive to $\alpha_1$ or $\alpha_2$, our assumptions regarding these 
parameters are not critical in any case.

Equation (6) relates $L_s(E,T)$ to $\Delta R/R$. Hence, measurement of $\Delta R/R$ at 
various current levels allows us to determine the spin diffusion length as a function of current 
through the organic. Knowing the current, we can find the electric field across the organic as follows: We apply Ohm's law to 
find the voltage $V$ across the organic from the relation $V = IR$, where $I$ is the current and $R$ is the measured resistance of the organic. 
The average electric field in the organic is then found from the relation $E = V/L$. This allows us to determine 
$L_s(E,T)$ as a function of $E$.
The electric field in the organic is of course not spatially uniform, but the arguments
presented here do not require the field to be uniform.

In Fig. 1, we plot  $L_s(E,T)$ versus $E$. The data show that the spin diffusion length $L_s(E,T)$ monotonically decreases 
with increasing electric field. 
That implies that the spin relaxation time decreases {\it very rapidly} with increasing electric field. To understand this, note first that 
we are operating in the high-field regime where $|E| \gg kT/\left ( e  L_s(E,T) \right )$. In our experiment, 
the average electric field strength $|E|$ varied between 3.16 kV/cm and 60 kV/cm, whereas $kT/\left ( e  L_s(E,T) \right )$
varied between 220 and 303 V/cm. This puts us in the high field regime. In this regime, Equations (1) and (2) yield 
\begin{eqnarray}
\left [ L_s(E,T) \right ]_{high~E} & \approx &  \mu(E, T) E \left 
[\tau_s(E, T) \right ]_{high~E} \nonumber \\
& = & \left \{ {{e d^2(E,T) \sqrt{E}}\over{\tau_0(E,T) \beta (T) }} tanh \left ( {{\beta (T) \sqrt{E}}\over{kT}} \right )
\exp \left ( {{\beta (T) \sqrt{E} - \Delta (T)}\over{kT}} \right ) \right \} 
\left [ \tau_s(E, T) \right ]_{high~E} \nonumber \\
& = & {{kT}\over{\beta}} \sqrt{E} \mu (0, T) tanh \left ( {{\beta (T) \sqrt{E}}\over{kT}} \right )
\exp \left ( {{\beta (T) \sqrt{E} - \Delta (T)}\over{kT}} \right ) \left 
[ \tau_s(E, T) \right ]_{high~E}, \nonumber \\
\label{relation}
\end{eqnarray}
which tells us that the spin relaxation time $\left [\tau_s(E, T) \right ]_{high~E}$
will have to drop off super-exponentially with the square-root of the average electric field in the organic if 
$\left [ L_s(E,T) \right ]_{high~E}$
 decreases with increasing field. Since that is the behavior of 
 $\left [ L_s(E,T) \right ]_{high~E}$ we observe experimentally, 
 we conclude that the spin relaxation time must have decreased very rapidly 
 with increasing average field in the organic.

The rapid decrease in the spin relaxation time with increasing electric field is once again {\it not} consistent 
with hyperfine interactions. To first order, the strength of hyperfine interaction 
is independent of the electric field. This strength is proportional to the sum of the carrier probability 
densities (squared modulus of the wavefunction) at the nuclear sites \cite{abragam}. 
An external electric field can skew the 
carrier wavefunctions in space (as in quantum confined Stark effect \cite{miller}) and change 
the interaction strength, but  it requires a very high electric field to
skew the wavefunction appreciably since  carriers in organics are quite strongly localized. Even in 
quantum confined Stark effect, where the carriers are relatively delocalized, it takes field strengths of several hundreds of 
kV/cm to change the overlap between electron and hole wavefunctions by a few percent. Therefore, 
we do not expect the hyperfine interaction strength to be
particularly sensitive to electric field. 

There is a second effect to be considered. Even though the 
hyperfine interaction strength may not be sensitive to electric field, the spin relaxation 
rate due to this interaction may become sensitive because the ensemble averaged spin relaxation rate 
$\left < 1/\tau_s \right >$ is equal to $\int_0^{\infty} d \epsilon f(\epsilon) \left [1/\tau_s(\epsilon) \right ]/\int_0^{\infty} d\epsilon f(\epsilon)$, 
where $f(\epsilon)$ is the carrier 
distribution function in energy space $\epsilon$. Since an electric field 
can change $f(\epsilon)$, it could influence $\left < 1/\tau_s \right >$, but such influence would be small because according 
to the variational principle of transport, a first order
change in the distribution function induces only a second order change in ensemble averaged 
transport parameters such as
$\left < 1/\tau_s \right >$ \cite{ziman}. Therefore, $\left [ \tau_s(E, T) 
\right ]_{high~E}$ could not be a strong function of electric field if hyperfine interactions were dominant. 

If $\left [ \tau_s(E, T) 
\right ]_{high~E}$ did not depend strongly on electric field -- as would be the case with 
hyperfine interactions -- then that would 
make the spin diffusion length $\left [ L_s(E,T) \right ]_{high~E}$ increase super-exponentially with the square-root of the electric field
according to Equation (\ref{relation}). This rapid increase in 
$\left [ L_s(E,T) \right ]_{high~E}$
with electric field is what ref. \cite{koopmans} also predicted if hyperfine interaction
is the primary spin relaxation mechanism. However, what we find experimentally is not a rapid increase but rather 
a {\it decrease} in  $\left [ L_s(E,T) \right ]_{high~E}$ with increasing field. This trend alone indicates that 
hyperfine interaction is most likely {\it not} the dominant spin relaxation mechanism in $Alq_3$.  Of course, hyperfine interaction is 
suppressed in a magnetic field \cite{koopmans, wohlgenannt},  and therefore it is possible that the magnetic fields used in the experiments of 
ref. \cite{pramanik}, quenched the hyperfine interaction. Nonetheless, we can say that at magnetic field strengths {\it commonly encountered 
in spintronic applications}, hyperfine interaction is not likely to be the major spin relaxing mechanism.

Finally, there are theoretical objections against hyperfine interaction as well. Most  organics of
interest are $\pi$-conjugated molecules where
the delocalized electron states are $p_z$ orbitals whose nodal planes coincide with the molecular plane.
Therefore, 
contact hyperfine interaction should be vanishingly small in them \cite{naber}. In some organic semiconductors like $Alq_3$, 
the electron wavefunctions may tend to localize over carbon atoms \cite{sanvito}, whose natural isotope $^{12}$C, has no net nuclear spin. 
Hence, contact hyperfine interaction should typically be weak in  organics.

Before we conclude, we point out that the data in Fig. 1 may actually suggest that the Elliott-Yafet mechanism \cite{elliott}
 is the major spin relaxer. This mechanism has its origin in the fact that any spin-orbit interaction makes the eigenspinors 
 of a carrier in the lowest unoccupied molecular orbital (LUMO) or highest occupied molecular orbital (HOMO)
  states of a $\pi$-conjugated molecule like $Alq_3$ momentum-dependent, so that whenever an electron or hole scatters and loses 
  (or gains) momentum, its spin relaxes. Although spin-orbit interaction in organics is  weak, it may not be so weak as to 
  preclude the Elliott-Yafet mechanism altogether. This is the conclusion we reached in ref. \cite{pramanik_nature} as well.

In order to understand why spin relaxation via the Elliott-Yafet mode could make 
$\left [ \tau_s(E, T) \right ]_{high~E}$
decrease rapidly with increasing electric field (and therefore make 
$\left [ L_s(E,T) \right ]_{high~E}$
decrease with increasing electric field -- consistent with the data of Fig. 1), consider the fact that 
in this mechanism, the spin relaxation rate 1/$\tau_s(E,T)$ is roughly proportional to the momentum relaxation rate 1/$\tau_m(E, T)$
and is given by \cite{yafet}
\begin{equation}
{{1}\over{\tau_s(E,T)}} \approx {{\Lambda(E)}\over{E_g}} {{1}\over{\tau_m(E,T)}},
\label{yafet}
\end{equation}
where $\Lambda(E)$ is the (electric-field-dependent) spin-orbit interaction strength in the LUMO levels for electrons or
 HOMO levels for holes, and $E_g$ is the HOMO-LUMO gap. The momentum relaxation rate 1/$\tau_m(E, T)$ will increase with electric field $E$
 because of enhanced scattering, but more importantly, the spin-orbit interaction strength $\Lambda(E)$ will also increase with electric field. 
 The spin-orbit interaction Hamiltonian is given by the expression  
\begin{equation}
H_{so} \propto  \left ( {\vec E} \times {\vec p} \right ) \cdot {\vec \sigma} ,
\end{equation}
where ${\vec p}$ is the momentum operator, ${\vec E}$
is the total electric field that the carrier sees (which includes the externally applied field) and ${\vec \sigma}$ 
is the Pauli spin matrix. 
Therefore,  $\Lambda(E)$ should increase with $E$. This is a well known fact in {\it inorganic}
 semiconductors and one of
its manifestation is the celebrated Rashba effect \cite{rashba}.
We can expect a similar effect in a disordered organic as well. That, taken together with the fact that the momentum
relaxation rate also increases with $E$, 
should make the spin relaxation time $\tau_s(E,T)$ decrease rapidly with increasing $E$. This
then should make $L_s(E, T)$ decrease with increasing electric field, consistent with our experimental observation
and the data in Fig. 1. 

The Elliott-Yafet mechanism is also consistent with the observed temperature dependence of the spin relaxation time in ref. 
\cite{pramanik_nature}. According to Equation (\ref{yafet}), the spin relaxation rate and the momentum 
relaxation rate should have the same temperature dependence since $\Lambda(E)$ and $E_g$ are nearly temperature 
independent. Hence 
$\tau_s(E,T)$ should exhibit weak temperature dependence if Coulomb scattering is the dominant momentum relaxing mechanism, since 
this scattering mechanism is elastic and makes the momentum relaxation rate nearly temperature-independent. In ref. \cite{pramanik_nature}, the major 
momentum relaxing mode was Coulomb scattering and $\tau_s(E,T)$ expectedly exhibited weak temperature dependence.
Therefore, in the end, both the temperature and the electric-field dependences of spin relaxation time are
consistent with the Elliott-Yafet mechanism, but not with hyperfine interactions.

In conclusion, we have shown that experimental evidence gathered so far and theoretical 
considerations tend to favor the Elliott-Yafet mechanism more than hyperfine interactions 
as the dominant spin relaxation mechanism of carriers in the most widely studied organics. 
Nonetheless, further experiments are required to resolve this issue conclusively.

%\pagebreak

\clearpage
\begin{figure}
\epsfig{file=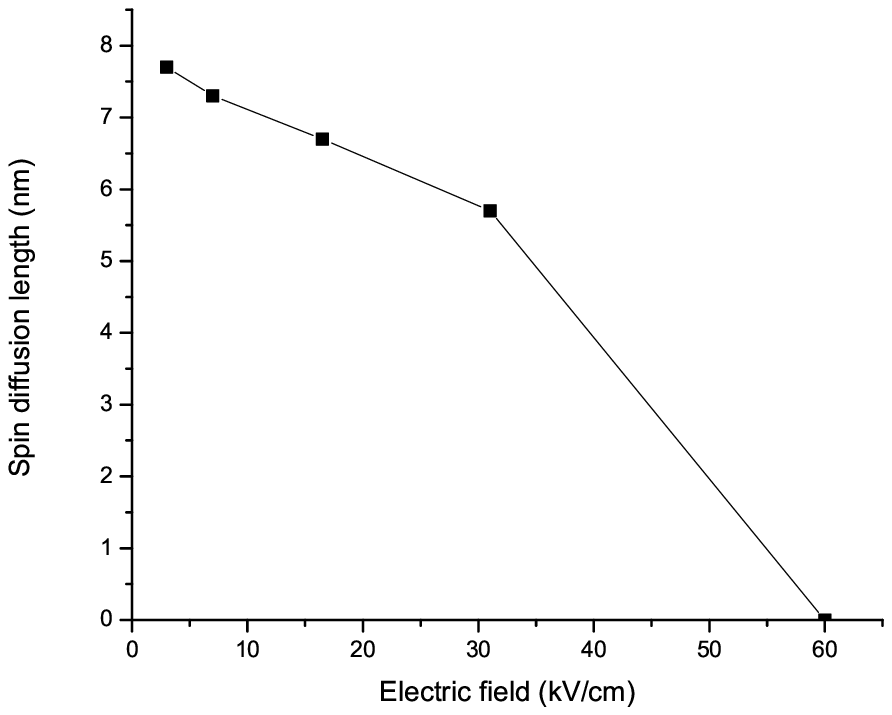, width=6.1in}
\caption{Spin diffusion length as a function of electric field. 
The measured spin diffusion length in 50-nm diameter $Alq_3$ nanowires as a function of electric field at a temperature of 1.9 K.
} 
\end{figure}

\end{document}